\documentclass[traditabstract]{aa} % for the abstract without structuration 
\usepackage{graphicx}
\usepackage{txfonts}
\usepackage{natbib}
\begin{document}
  \title{High Precision Radial Velocity Measurements in the Infrared\footnote{
based on observations at the European Southern Observatory,
Paranal, during commissioning of CRIRES.}}
    \subtitle{A First Assessment of the RV Stability of CRIRES}

   \titlerunning{Radial Velocity stability of CRIRES}

   \author{A. Seifahrt
          \inst{1,2,3}
          \and
           H. U. K\"{a}ufl\inst{3}}

   \institute{Universit\"at G\"ottingen, Institut f\"ur Astrophysik, Friedrich-Hund-Platz 1, D-37077 G\"ottingen, Germany\\
     \email{seifahrt@astro.physik.uni-goettingen.de}
     \and
     Universit\"at Jena, Astrophysikalisches Institut und Universit\"ats-Sternwarte, Schillerg\"asschen 2, D-07745 Jena, Germany
     \and
     ESO, Karl-Schwarzschild-Str. 2, D-85748 Garching, Germany}

   \date{as of \today}

% \abstract{}{}{}{}{} 
% 5 {} token are mandatory
 
  \abstract
{ High precision radial velocity (RV) measurements in the near infrared are on
  high demand, especially in the context of exoplanet search campaigns
  shifting their interest to late type stars in order to detect planets with
  ever lower mass or targeting embedded pre-main-sequence objects.

ESO is offering a new spectrograph at the VLT -- CRIRES -- designed for high
resolution near-infrared spectroscopy with a comparably broad wavelength
coverage and the possibility to use gas-cells to provide a stable RV zero-point.

We investigate here the intrinsic short-term RV stability of CRIRES, 
both with gas-cell calibration data and on-sky measurements using the absorption 
lines of the Earth's atmosphere imprinted in the source spectrum as a local RV
rest frame. Moreover, we also investigate for the first time the intrinsic
stability of telluric lines at 4100\,nm for features originating in the 
lower troposphere.

Our analysis of nearly 5 hours of consecutive observations of MS~Vel, a M2II
bright giant centred at two SiO first overtone band-heads at 4100\,nm, 
demonstrates that the intrinsic short-term stability of CRIRES is 
very high, showing only a slow and fully compensateable drift of up to 60\,m/s 
after 4.5 hours. The radial velocity of the telluric lines is constant down to 
a level of approx. $\pm10$\,m/s (or 7/1000 of one pixel). Utilising the same telluric
lines as a rest frame for our radial velocity measurements of the science target, 
we obtain a constant RV with a precision of approx. $\pm20$\,m/s for MS~Vel as expected
for a M-giant.}


   \maketitle
%
%________________________________________________________________

\section{Introduction}

CRIRES\footnote{CRIRES stands for \underline{CR}yogenic \underline{I}nfra\underline{R}ed
    \underline{E}chelle \underline{S}pectrograph} is a newly commissioned adaptive optics (AO) - fed spectrograph at UT1
(Antu) of the Very Large Telescope on Paranal, Chile. The spectrograph provides 
a resolution of up to 100\,000 at 960--5200\,nm with an instantaneous wavelength 
coverage of $\sim\lambda/50$ on a 4096$\times$512 pixel detector mosaic \citep{kaufl04,
  kaufl06b}.  %%%% new reference
A MACAO\footnote{MACAO stands for \underline{M}ultiple \underline{A}pplication
    \underline{C}urvature \underline{A}daptive
    \underline{O}ptics} %%%% new footnote
adaptive optics system \citep{arsen2004}
%%%% new reference
is providing diffraction limited images at the entrance slit of CRIRES with
Strehl ratios of up to 65\% in the K band, thus roughly doubling the
throughput under reasonable seeing conditions for the nominal 0.2\arcsec~slit
when compared to seeing limited performance \citep{jerome06}.

The spectrum, being either a full order at the shortest wavelength in the J
band and about 1/5 of an order at the longest wavelength in M band, is imaged
in long slit mode onto a mosaic of four Aladdin III detectors with a gap of
$\sim 250$ pix between the chips. Thus, the instantaneous wavelength coverage
is much shorter than for optical cross-dispersed spectrographs.  However, the 
gain in efficiency for RV measurements, primarily dependent on spectral coverage 
and S/N, will shift in favour of CRIRES as soon as the effective SED of the science
target peaks longwards of 900\,nm where other optical spectrographs commonly
used for RV measurements, are either not operating 
(e.g. HARPS) or loose dramatically in sensitivity (e.g. UVES).

For precise RV measurements the spectrograph has to be well
characterised since we deal with various sources of uncertainties that could
induce wavelength shifts in the measured spectra.  Since CRIRES is a
cryogenic, thus actively stabilised spectrograph, temperature
drifts are small but certainly not negligible. The same is true for vibrations
and flexure, especially due to changing thermal gradients in the structure. 
Being mounted on a Nasmyth platform of the VLT, the system is intrinsically
very stable but vibrations induced by the cryogenic cooling system can be
suspected for causing instability.

A second effect is the imperfect repeatability of dispersive parts on moving
functions suffering slip-stick effects: the prism (pre-disperser), the
intermediate slit and the grating. It is forseen that functions are left
untouched when the same instrumental setup is used in subsequent observations.  
However, for observations that span days to months the repeatability of the 
setup could be an issue, if it can't be well calibrated.

Last but not least a long-slit spectrograph suffers the natural problem of
alignment uncertainties of the source with respect to the slit centre. For
spectrographs fed by an adaptive optics system the situation is worse in some
aspects:\label{DCR}
\begin{enumerate}
 \item The FWHM of the source is usually smaller than the slit width.
 \item The alignment of the adaptive optics in respect to the entrance slit of
   the spectrograph suffers instabilities, requiring an active slit-viewer
   guiding in order to keep the source centred on the slit.
 \item The algorithm for slit-viewer guiding can only work on the wings of the
   stellar PSF, since its core is masked by the slit.
 \item The refraction caused by the Earth's atmosphere has to be measured and
   compensated simultaneously as observations are performed at three different
   wavelengths: in the optical for the AO wave front sensor, 
   in J, H or K band for the slit-viewer (used for active
   guiding) and at the respective wavelength of the observation, in our case
   at 4100\,nm.
\end{enumerate}
%%%% new
All these effects will cause a degraded stability and a possible drift of the
wavelength scale with time. We thus aimed to characterise the system and the
means to calibrate it and test the on-sky performance in terms
of RV stability during a single night.

For CRIRES a RV precision of 15~m/s would require pointing and tracking with a
precision of one milli~arcsec (mas), given that one pixel in dispersion 
direction (1.5 km/s) equals 100~mas at the entrance slit of the spectrograph. 
The tracking precision is limited by the amount of light in the wings of the PSF 
on which the guiding algorithm of the slitviewer operates. Tracking the core of 
the PSF on the seeing-limited wings is a general limitation. Even for bright sources, 
the performance of the guiding algorithm is not sufficient to stabilise the source 
in the slit with a precision better than approx. 25 mas (approx. 200~m/s), which 
makes a reference source (telluric or gascell absorption lines) indispensable 
when aiming for high precision RV measurements.

\section{Observations}
For our on-sky test we were looking for a bright target which should present
no significant RV signal within a few hours. 
We chose the M2II star MS~Vel
\citep{egret80}, a rather bright ($Ks = 3.7$ mag (DENIS), $F_{4.29\mu m} = 16$
Jy \citep{egan03}) semi-regular variable star \citep{adelman01} with a
proposed period of about 360d \citep{makarov04}.  We have re-analysed
HIPPARCOS and TYCHO photometry simultaneously covering about 3 years and about
5.5 years worth of data from the ASAS3 V band survey \citep{asas} using a
Lomb-Scargle Periodogram to identify variability in MS~Vel. All three data
sets show a strong period of $P=360 \pm 5$d with a somewhat larger value ($P
\sim 363.5$ d) for the HIPPARCOS and TYCHO epoch of 1990 - 1993 and a shorter
value ($P \sim 355$ d) in the full epoch covered by the ASAS data (2000 -
2006). This dominant period with an amplitude of about $\Delta V \simeq 0.8$
mag is accompanied by shorter periods of 18, 26 and 28 days with much smaller
amplitudes in the earlier data which disappear however in the much better
sampled data from the later epoch. Conversely, periods of 155 and 185 days,
present in the ASAS data can not be found in the HIPPARCOS and TYCHO data. All
these periods have false-alarm-probabilities of much less than 0.1\% and are
therefore real, even though not stable. We see that MS~Vel is variable on
various time-scales of weeks to a about a year. However, it is very unlikely
that the object shows a significant RV signal in only a few hours, given its
large radius. As predicted by 3D simulations \citep{Freytag08}, typical 
photometric periods for early to mid M-type giants are at the order of a 
several dozend to a few hundred days \citep{Percy08}. Radial velocity surveys 
of K-type giants also revealed periods at the same timescale \citep{Hekker08}. 
Short-term oscillations on M-type giants are rarely observed and still show
periods in excess of one day \citep{Koen00}.

We have observed MS~Vel in the night of 2007, Feb 02 from 03:23 to 07:56
UT. Observations were obtained with the adaptive optics loop closed and locked on 
the science target. Strehl ratios of more than 50\% could be achieved during the 
majority of the program, delivering most of the encircled energy in the diffraction limited
core with a FWHM of approx. 100~mas. The DIMM seeing was 1\arcsec\, on average, with very 
good conditions (0.65\arcsec--1.0\arcsec\,) for the first three hours and slightly worse 
conditions (1.0\arcsec--1.5\arcsec\,) for the last 90~min. A slit width of 
0.4\arcsec\, was chosen, providing a nominal resolution of 
R$\simeq$ 50\,000. The total flux in the spectrum shows a weak dependence on the 
seeing, with a degradation of about 20--30\% in flux at the end of the campaign, 
as expected when the atmospheric conditions worsens and the Strehl ratio drops, 
matching the predictions of the exposure time calculator (ETC) for our case very nicely.

We chose a setting at $\lambda_{Ref}=4029.9$ nm to cover two SiO band-heads at 4043 nm 
($^{28}$Si$^{16}$O, $\nu$=3$\rightarrow$1) and 4084 nm ($^{28}$Si$^{16}$O, $\nu$=4$\rightarrow$2). 
The range of the SiO-overtone at 4000--4100nm is a well understood but still rich line system  
\citep{aringer1418} in a region where photospheres are otherwise generally clean while the 
Earth's atmosphere provides for a rich line system of N$_2$O, originating mainly from 
the lower troposphere.

An AAA-BBB-BBB-AAA nodding pattern with individual integration times of 10\,s per 
frame was used. The nodding amplitude was set to 10". The average S/N in each single 
spectrum was found to be approx. 200.

After each cycle the instrument 
switched to calibration mode to take a single, short spectrum of the N$_2$O gas-cell
(10~hPa, 140~mm length),
illuminated by a blackbody involving an Ulbricht sphere. The slit width was
held constant for the science and calibration measurements. Since the light from the
gas-cell is a spatially flat source, the entrance slit of CRIRES is evenly illuminated
and the 0.4\arcsec\, slit delivers a resolution of approx. 50\,000. No change in
illumination is expected throughout the length of the program, making these calibration
measurements a suitable reference for the intrinsic stability of the spectrograph, 
independent from slit centring issues of the actual science target. 
After completing such a set of observations we re-acquired the object in the slit-viewer, 
closed the AO loop again and started a new cycle of on-sky observations. This procedure
was repeated 25 times.

During this series of observations, the dispersive elements of CRIRES -- the
prism (pre-disperser), intermediate slit and the grating -- where electronically
de-mobilised and held in fixed positions. This precaution was taken to avoid
accidental repositioning of functions. Thus, the observations of the calibration
source were only affected by temperature drifts and flexure. Besides these non standard
setup the observations made use of ESO observatory standard procedures.

Both, the signal-to-noise ratio and the spectral resolution are of critical 
importance for RV studies. While both quantities are closely connected to the
slit width of regular long-slit spectrographs, the situation for CRIRES is more
complex. Since CRIRES works with an underfilled entrance slit, the 
achievable resolution is not strictly depending on the chosen slit width but 
is mainly governed by the FWHM of the stellar PSF in the slit, hence, also 
depends on the AO performance and the wavelength of the observation. The troughput is 
only marginally dependant on the slit width in the regime of potentially 
high Strehl ratios in $K$, $L$, and $M$ band, reaching over 80\% at the longest
wavelength under favourable conditions \citep[see the CRIRES ETC and ][]{jerome06}. 
The typical diffraction limited FWHM of the VLT at 4000\,nm ($\sim$0.1\arcsec\,) 
is much smaller than the minimum slit width and the spectral resolution is limited 
by the optics of the spectrograph and the stability of the source in the slit during 
an integration rather than by the width of the profile alone or the width of the 
entrance slit. The true resolution achieved with the 0.4\arcsec\, slit is thus 
likely closer to 100\,000 than to its lower limit of 50\,000, defined by the slit 
width. Closing the slit to 0.2\arcsec\, would only marginally improve the resolution 
by basically masking some light from the seeing limited halo of the PSF. Likewise, 
the troughput of the 0.4\arcsec\, slit is only about 20\% higher than for 0.2\arcsec\, 
slit in our case.

However, choosing a wide slit further decreases the amount of reflected light available for
the guiding algorithm of the slitviewer and further increases the chance of
misplacements and instabilities of the source position in the slit. Since 
it is expected that future RV programs conducted with CRIRES will also observe 
rather faint targets under low Strehl ratios, a regime where a large slit width 
will more notably improve the efficiency, we nevertheless decided to use the 
0.4\arcsec\, slit to investigate these effects also in this first study.

\section{Data Reduction}

\subsection{Basic Treatment and Spectral Extraction}
For data reduction the raw data were divided by a flat-field frame, mainly to
correct for a fixed pattern effect of the detector array and to normalise the
continuum to a satisfactory level. To subtract the background emission we
subtracted the mean of the three frames from the next nodding position from 
the current frame, leaving a flat sky background around the source spectrum.
The spectra were than extracted using the optimum extraction
algorithm of the CRIRES data reduction pipeline.

Spatially extended sources, such as the sky background as well as the spectra
of the gas-cell in the respective calibration frames have been extracted after 
flat-fielding of the raw frames and integrating over a spatial range of 100~pixels 
between the two nodding positions of the source. We thus sample the same region that
is covered by the source spectrum itself, without introducing offsets caused
by slit curvature, detector residual tilt, and distortion, as we will discuss
in Sect.~\ref{here}.

After this treatment we have 300 individual spectra of the
object and of the sky background emission at our disposal. In addition we have 
25 spectra, one for each cycle of observations, of the N$_2$O gas-cell. Hence, 
we have four systems of spectral lines that we can analyse here:

\begin{enumerate}
\item the photospheric lines of the target, MS~Vel, foremost the two SiO
  band-heads and various unidentified lines,
\item the telluric lines in absorption, imprinted in the spectra of the
  target,
\item the telluric lines in emission as a spatially extended background source
  along the whole slit (note: due to Kirchhoff's law the atmospheric
  absorption lines become inverted into emission lines when viewed against the
  night sky) and finally
%%%% ich hab den Kirchhoff hier eingebaut, damit die das auch verstehen
\item the N$_2$O lines from the gas-cell, also as a spatially extended source
  since the instruments flat-field source was used to illuminate the gas-cell.
\end{enumerate}

All four line systems can now be analysed by means of cross-correlation to
measure the zero-point shift of the wavelength in each of the spectra.

\subsection{Methodology}
\begin{figure*}
\centering
\includegraphics[width=17cm]{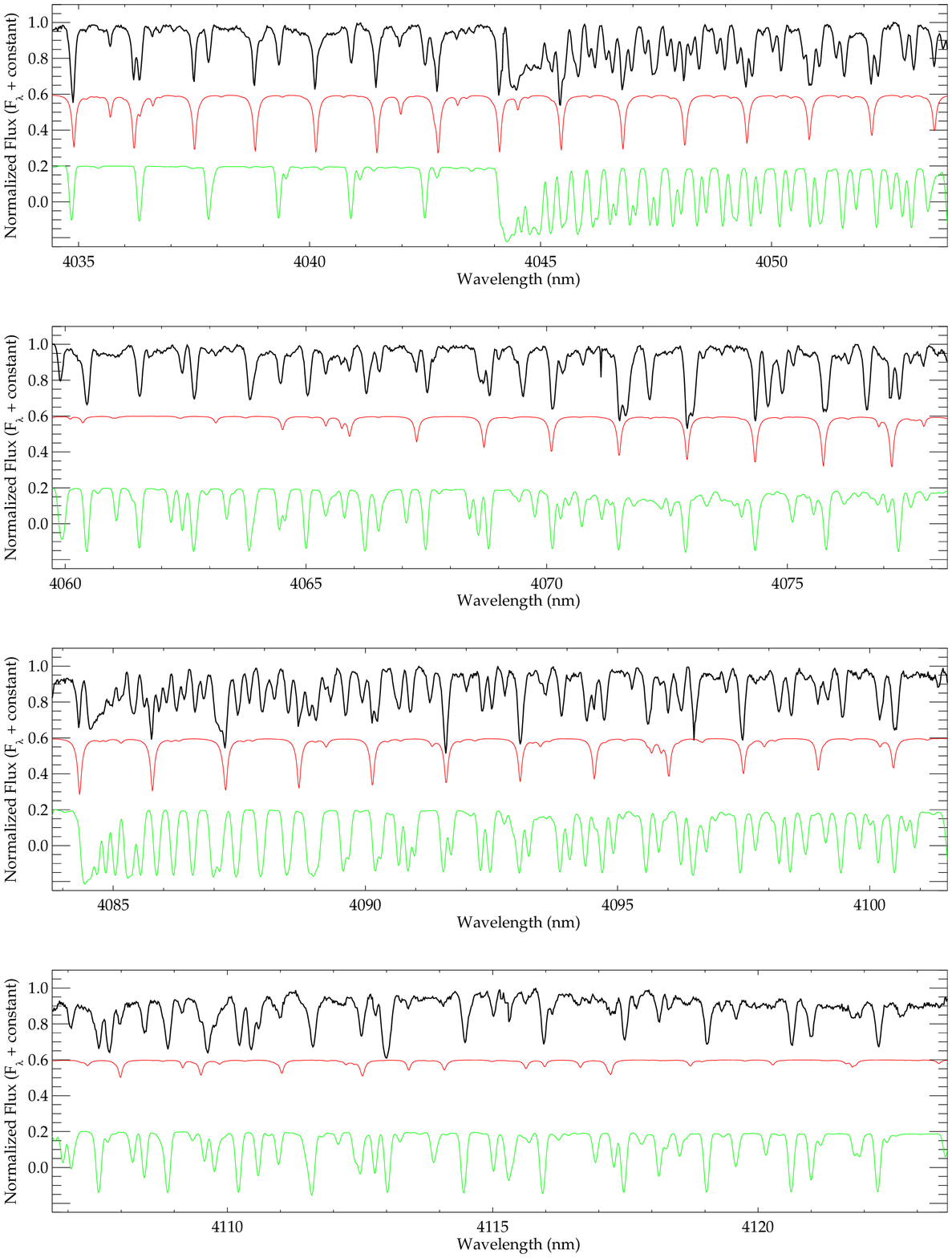}
\caption{Spectrum of MS~Vel and the Earth's atmosphere on the four chips of CRIRES, 
displayed in four viewgraphs. Shown in each viewgraph: measured spectrum 
(top, black line), telluric model (center, red line,shifted by -0.4) and PHOENIX model 
spectrum for MS~Vel: T$_{eff}$=3500\,K and $\log{g}$=0.5 (bottom, green line, shifted 
by -0.8). See text for details.
\label{fig:sourcespectrum}}
\end{figure*}

In case the spectrum of the star is taken through a gas-cell, the classical
approach to measure radial velocities is to cross-correlate the product of two
template spectra (normally the stellar spectrum and the spectrum of the
gas-cell) with the actual measurement. This requires template spectra of high
quality. Unfortunately no template spectra for a M2II giant at
4100\,nm of sufficient quality was available for us. A test with a synthetic
spectrum from the library of PHOENIX \citep[][and references therein]{Hauschildt} spectra\footnote{kindly provided by Peter
Hauschildt, Hamburg}, as displayed in Fig.~\ref{fig:sourcespectrum}, did not
lead to consistent results when used as a template for the cross-correlation
analysis, since the line lists used to compute the synthetic spectra are not 
precise enough and show notable shifts of the different species present 
in late type giants at 4100\,nm.

\begin{figure*}
\centering
\includegraphics[width=17cm]{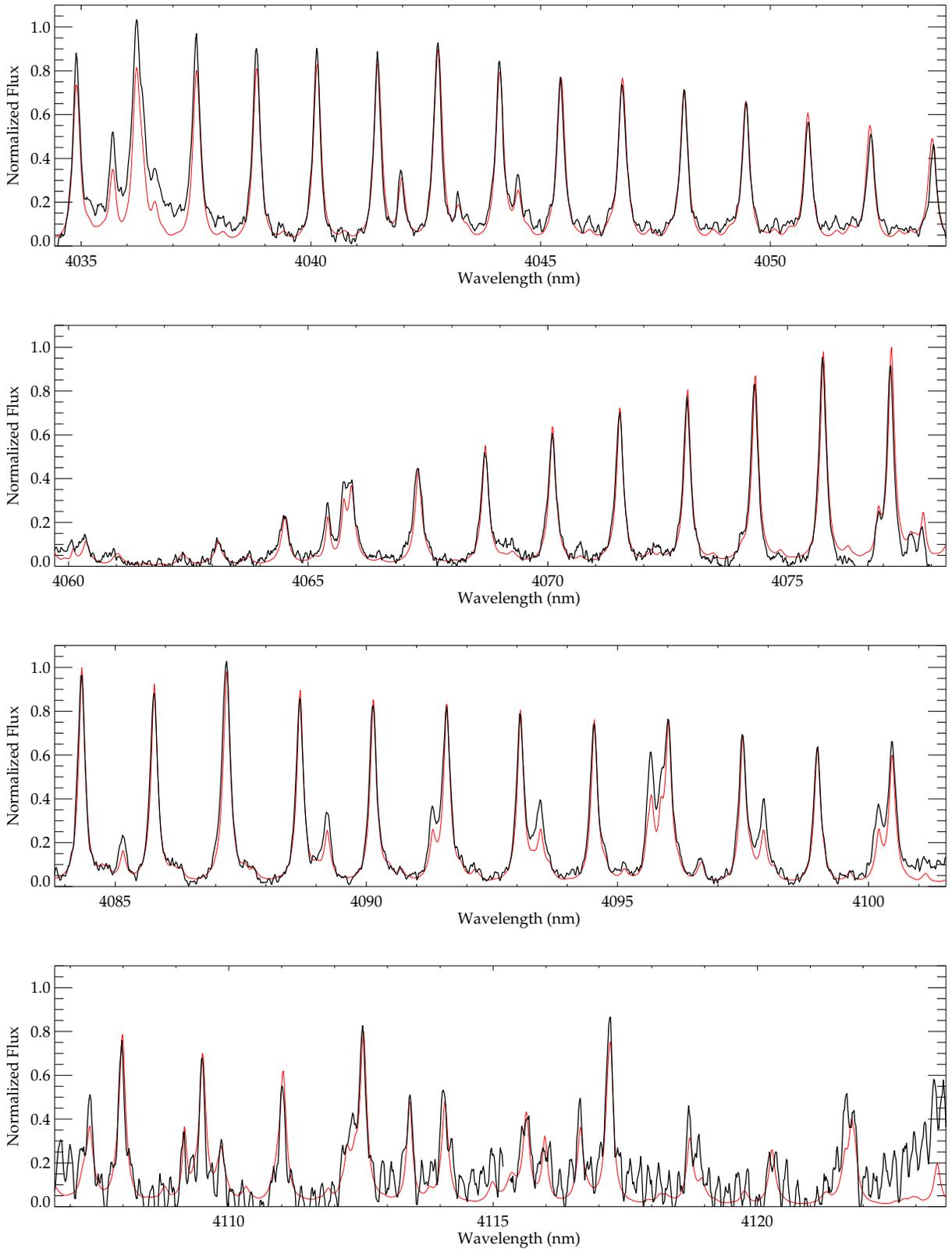}
\caption{Spectrum of the Earth's atmosphere in emission (sky background) on the four chips of CRIRES, 
displayed in four viewgraphs. The relative strength of the emission bands follows the behavior of the 
N$_2$O absorption bands shown in Fig.~\ref{fig:gascellspectrum}. The relative intensities on chips 2, 3, and 4
are scaled up by factors of 7\%, 15\% and 150\%, respectively, relative to chip 1. Over-plotted is a 
synthetic emission spectrum (red line). Note that the isolated H$_2$O lines appear much stronger in emission
than they appear in absoption in Fig.~\ref{fig:sourcespectrum}. See text for further details.
\label{fig:skyspectrum}}
\end{figure*}

For the telluric lines however, we have access to quite precise synthetic
spectra of the Earth's atmosphere. Using a standard atmospheric model for the
Paranal site and adopting a typical humidity for the time of the observation
we can construct a synthetic spectrum at any given wavelength using the
FASCODE algorithm \citep{Clough81,Clough92}, a line-by-line radiative 
transfer model for the Earth's atmosphere, and HITRAN \citep{HITRAN2004}
as a database for molecular transitions. A spectrum of the Earth's atmosphere 
in our wavelength window is shown in Fig.~\ref{fig:sourcespectrum} for 
the telluric lines in absorption as seen against the stellar continuum. 
In Fig.~\ref{fig:skyspectrum} we show the same lines in emission (line 
radiation against dark sky background). It is worth noticing that the predominant
species in the observed spectral range is in fact N$_2$O, the same species as
in our gas-cell. N$_2$O shows nearly regular spaced lines of the $P$-branch on 
chips two, three, and four and of the $R$-branch on chip one. Additional minor 
contributions arise from some isolated water vapour lines. These lines form in lower, 
thus warmer parts of the atmosphere and are stronger in emission than in absorption 
when compared to the neighboring N$_2$O lines. They are thus more prominent in the 
sky emission spectrum (Fig.~\ref{fig:skyspectrum}) than in its 'inverse' counterpart, 
the telluric absorption spectrum (Fig.~\ref{fig:sourcespectrum}).

A more detailed discussion of the properties of the telluric lines and the performance
of using them as wavelength standards will be given in a later paper.

\begin{figure*}
\centering
\includegraphics[width=17cm]{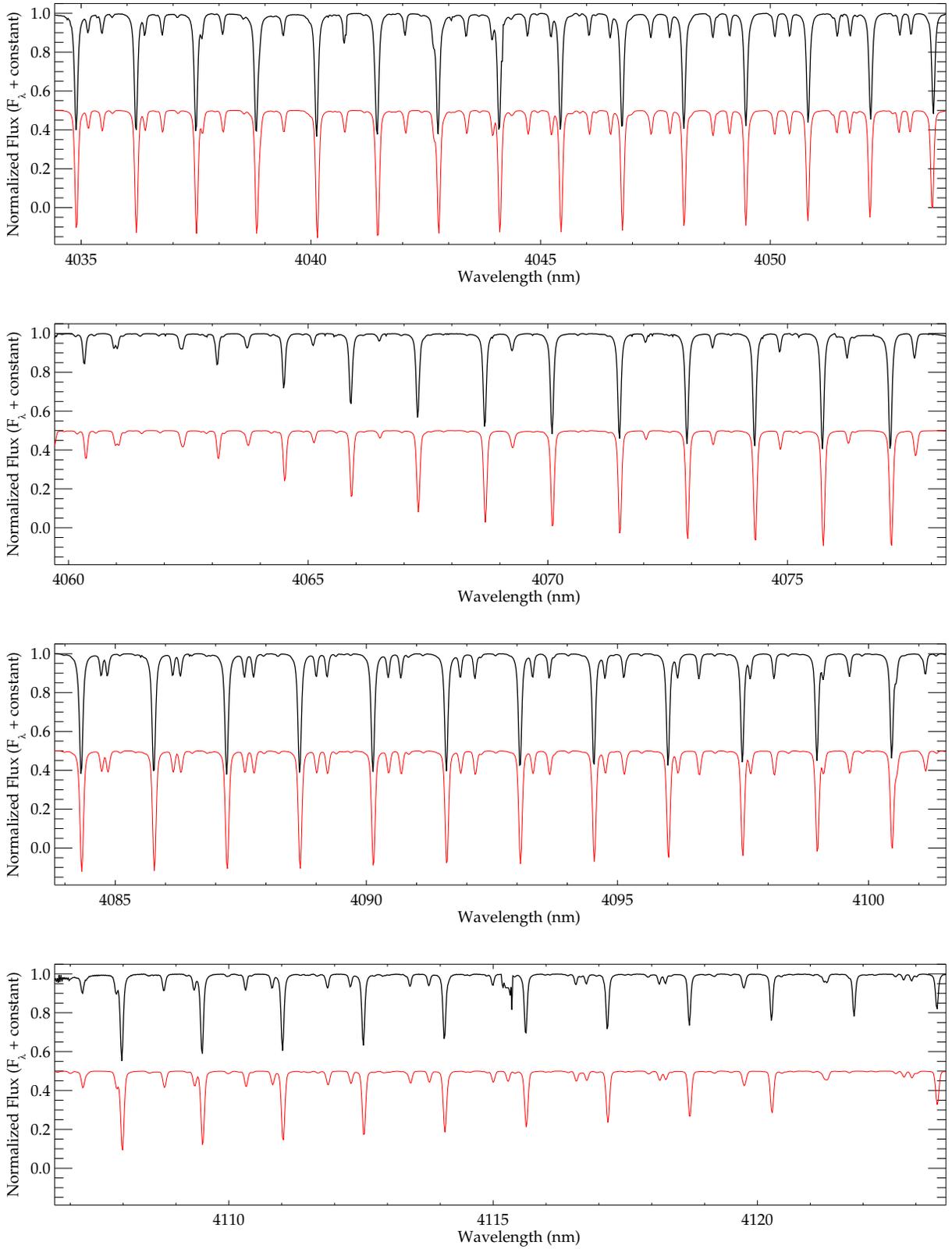}
\caption{Spectrum of the N$_2$O gas-cell, illuminated by the flat-field 
lamp in calibration mode. 
Overplotted is a synthetic spectrum (red line, offset -0.6), calculated with FASCODE and HITRAN.
See text for details.
\label{fig:gascellspectrum}}
\end{figure*}

Similarly to the telluric features, we can compute the spectrum of our gas-cell, 
a N$_2$O spectrum at room temperature and $\sim$10\,hPa pressure.  The result is shown in
Fig.~\ref{fig:gascellspectrum}. These transitions are calibrated in frequency
with heterodyne techniques against the time standard \citep{nist1992}, so that
they provide for an absolute precision, traceable to the time standard, not yet
available for optical spectroscopy \citep{li2008}. 

\subsection{Wavelength Calibration}
For a RV measurement we need to determine the wavelength of each pixel to a
high precision.  The gas-cell delivers a N$_2$O absorption spectrum of high S/N
with nearly evenly spaced lines being distributed in high density on all four
chips (see Fig.~\ref{fig:gascellspectrum}). Using the fore-mentioned synthetic
spectrum of our gas-cell, we fitted a quadratic solution independently to each
of the four chips to match the measured spectrum with the synthetic spectrum,
thus providing the zero point and dispersion for the four chips in the first 
recorded N$_2$O spectrum during our measurements.

\subsection{Cross-Correlation}
Given the lack of a proper template spectrum for the science target, we use
the following approach to determine the radial velocity of each of the line
systems as outlined above: (1) For the photospheric lines of MS~Vel we mask
the telluric lines for transmissions below 95\% and cross-correlate each of the
300 spectra against the first one obtained. (2) The wavelength shifts of all
other line systems -- the telluric absorption lines in the spectrum of MS~Vel,
the sky-background emission lines and the N$_2$O absorption lines -- are
computed by cross-correlating our measured spectra against the respective
synthetic model spectrum. 

We also tested the methodology used for the photospheric lines of MS~Vel 
(cross-correlation of all spectra against the first one) for the N$_2$O
lines and the telluric lines. We obtained the same results as for the 
cross-correlation with the synthetic templates, but find an enhanced scatter
due to the lower signal-to-noise when cross-correlating to a measured spectrum
instead of against a noise-free model.

\section{Results and Discussion}

\begin{figure*}
\centering
\includegraphics[angle=90,width=17cm]{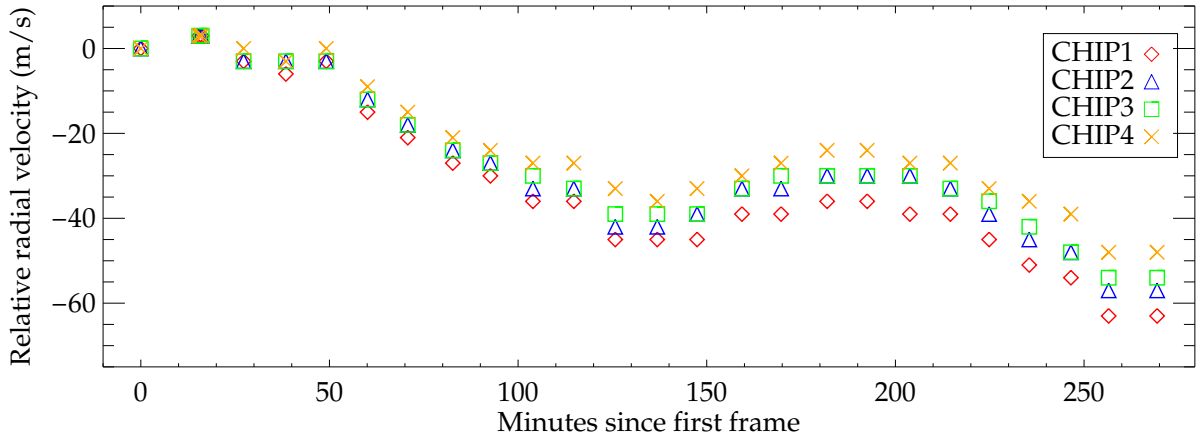}
\caption{Apparent radial velocity of the N$_2$O gas-cell line system 
(see Fig.~\ref{fig:gascellspectrum}; displayed separately for 
each chip) observed in calibaration mode (the gas-cell being illuminated by the 
flat field lamp). Note the change in dispersion with time which manifests as 
divergent trends in the RV measured on the four chips. See text for details.
%%%% comb bashing still missing
\label{fig:RVgascell}}
\end{figure*}

The 25 calibration spectra of the N$_2$O gas-cell are distributed in time over
the full length of our experiment.  These spectra probe the intrinsic
stability of the spectrograph, since the illumination of the gas-cell (and thus
of the entrance slit) is spatially flat and we do not trace any slit
alignment effects. Moreover, the temperature of the gas-cell, even though not
actively stabilised, is certainly not variable to a level that could induce line
broadening and line-shifts in the N$_2$O spectrum\footnote{ During our
  observations the external temperature as logged by the observatory dropped
  from 14.8 to 13.3~C. The gas-cell has some thermal inertia, but will
  ultimately follow the external temperature. Assuming a typical pressure
  induced  frequency shift of 100-150~MHz for 1000~hPa \citep{Deming}, even the 
  maximum temperature change for Paranal ($-5\, \leq \, T_{amb}\, \leq 20\,C$) 
  will not result in a shift of more than 150~kHz of the line frequencies, 
  equivalent to a velocity error of 50~cm/s.}. 
%%%% die footnote ist neu, sowie die meteo daten
Hence, we adopt
this spectrum as a true rest frame. As can be seen in Fig.~\ref{fig:RVgascell},
the spectral lines of the gas-cell show a slow drift in wavelength that
culminates in a RV offset of $\sim$ 60\,m/s after 4.5\,hours. The drift is smooth
but non-monotonous and is very similar for all four chips leading only to a
minimal degree of dispersion change, as can be seen by the slightly diverging
RV trends of the four chips. Such a change in dispersion can be assigned to a 
small drift in the temperature of the grating. At the time of our measurements,
the grating was only passively stabilised and the temperature sensor at the grating
had a limited resolution of only $\sim0.1$~K. Moreover, the sensor was not part of 
an active temperature stabilisation of the grating. Following the basic grating equation 
for echelle spectrographs, the angular dispersion changes linearly with the groove spacing of 
the grating (induced e.g. by thermal expansion of the grating material) when the setup 
is otherwise held constant. Adopting a typical expansion coefficient of $dl/l \simeq 
4\cdot10^{-6}$~K$^{-1}$ for the grating material, a temperature drift of only 50~mK over 
4.5 hours could have caused the observed drift of 60\,m/s. Such a drift would have easily 
passed the sensor unnoticed. This shortcoming has been fixed by now. The grating is now 
actively stabilised to 1~mK. The expected wavelength stability is thus approx. 1~m/s.

This drift would be seen in any spectral line that is intrinsically stable and
used as a local rest frame and can thus be compensated for. Ideally the gas-cell
should be used simultaneously while measuring the spectra of the science
target, the absorption lines of the gas being imprinted into the source
spectrum. However, the chosen gas has to match the spectrum of the object and
the gas should not be present in the Earth's atmosphere to avoid any overlap
or mixing of the spectrum of the Earth's atmosphere and the spectrum of the
gas-cell. At the time of our experiment we had only one gas-cell ready and its
content, N$_2$O, is also highly abundant in our atmosphere.  Hence, we decided
to use the gas-cell only in calibration mode, illuminated by the flat-field
source, to trace the internal stability of the spectrograph and use the
atmospheric lines, seen in absorption in the target spectrum, as our local
rest frame.

\begin{figure*}
\centering
\includegraphics[angle=90,width=17cm]{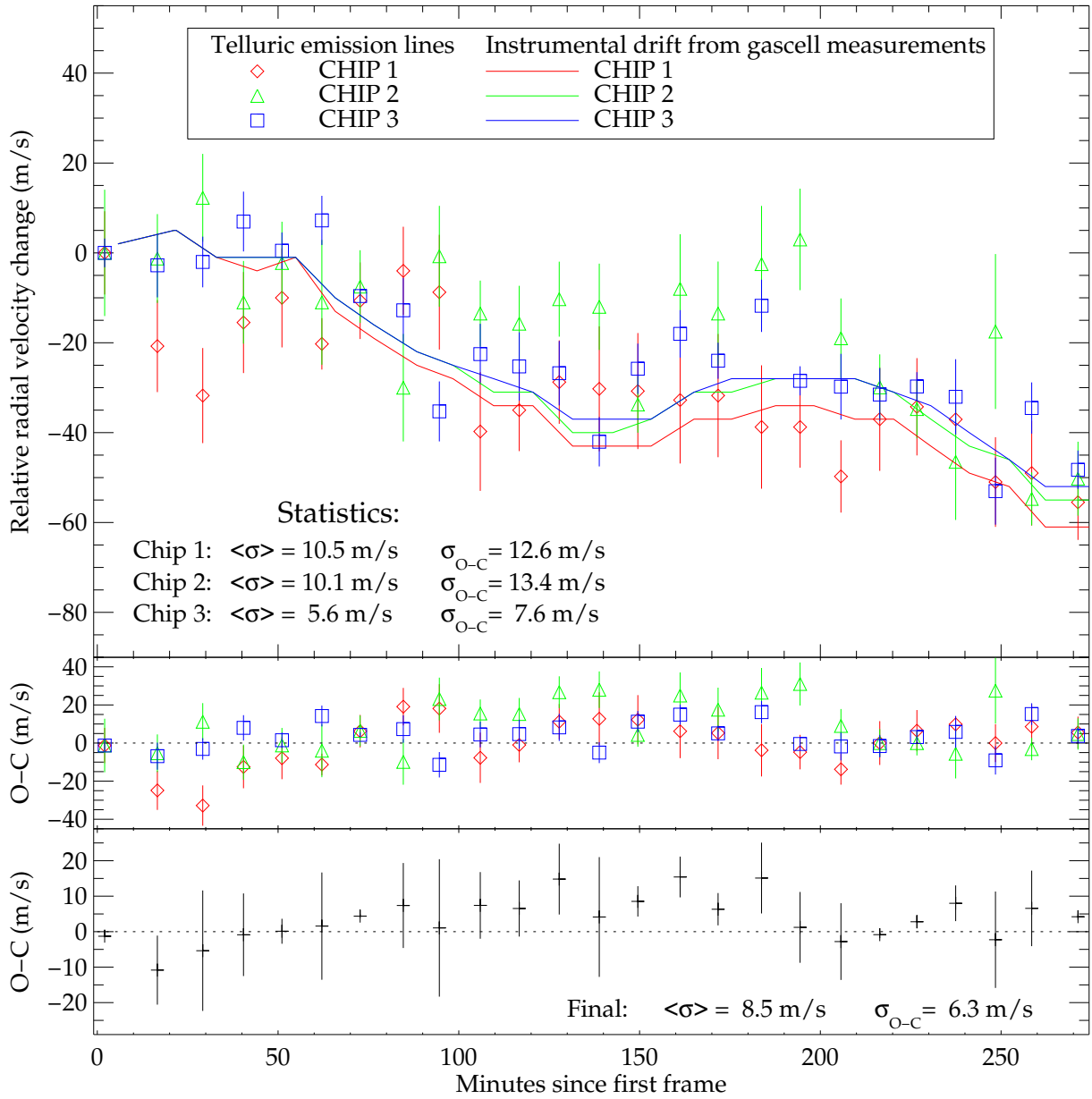}
\caption{Radial velocity of the telluric emission lines (see Fig.~\ref{fig:skyspectrum}) and the N$_2$O gas-cell line system (see Fig.~\ref{fig:gascellspectrum}) showing the instrumental drift during the measurement. In the subpanels the difference between both measurements is shown in terms of observed - computed (O-C), first separately for each of the three chips analyzed here, finally for the mean of all three chips. The standard deviation of the mean was used to derive formal error bars. Residuals are the final uncertainty for the RV stability of the telluric lines. See text for details.
\label{fig:RVemission}}
\end{figure*}

Conversely, we can use the atmospheric lines in emission to trace the behaviour
of these lines - to check for their stability. Here the same argument holds as
for the N$_2$O lines: The atmospheric lines in emission are a spatially flat
source unaffected by any slit misalignments. 
%%%% new
The frequency of the atmospheric lines can be systematically shifted if there
were a net radial velocity introduced by wind or turbulence.
%%%%
Moreover, these lines will be affected by the slow drift of the spectrograph
itself, measured with the N$_2$O gas-cell lines. In Fig.~\ref{fig:RVemission} we show
the RV trend of the atmospheric lines in emission. The measured RVs from 12
spectra taken within one nodding cycle are averaged to a single data point. The
standard deviation of the mean is shown as the formal 1$\sigma$ error of each
data point. We over-plot the drift of the spectrograph and subtract this drift
from the RV points. Combining the results from all three chips where
atmospheric lines are present results in a flat RV distribution over time with
residuals (r.m.s.) in the order of $\leq$ 10\,m/s which is also well within the formal
uncertainties derived for each data point.  Thus, we conclude at this point that
the atmospheric lines are well suited as a reference to trace the RV of the
science target. We also note here that the spectral resolution of the emission
lines is limited by the slit width to R$\sim$50\,000 while the resolution of
the object spectrum is more dependent on the true FWHM of the PSF in the slit,
thus providing a higher resolution.\label{here}

Finally we measure the RV of the photospheric lines of MS~Vel and in parallel
the atmospheric lines in absorption, as outlined above, that is by
cross-correlating the masked spectra of the source against the first measured
spectrum and the unmasked spectra against a template of the atmospheric
absorption lines.  Both line-systems show a strong near-linear trend with time, 
peaking at $\sim$ 1.5\,km/s after 4.5 hours. This trend is most likely caused
by residuals from the correction of the chromatic refraction (see point 4 in 
Sect.~\ref{DCR}.) In addition, the nodding pattern induces sudden jumps of up 
to 500\,m/s (1/3 pixel) between the two nodding positions due to the
slit curvature and a slight rotation of chips one and two in the focal plane. 

\begin{figure*}
\centering
\includegraphics[angle=90,width=17cm]{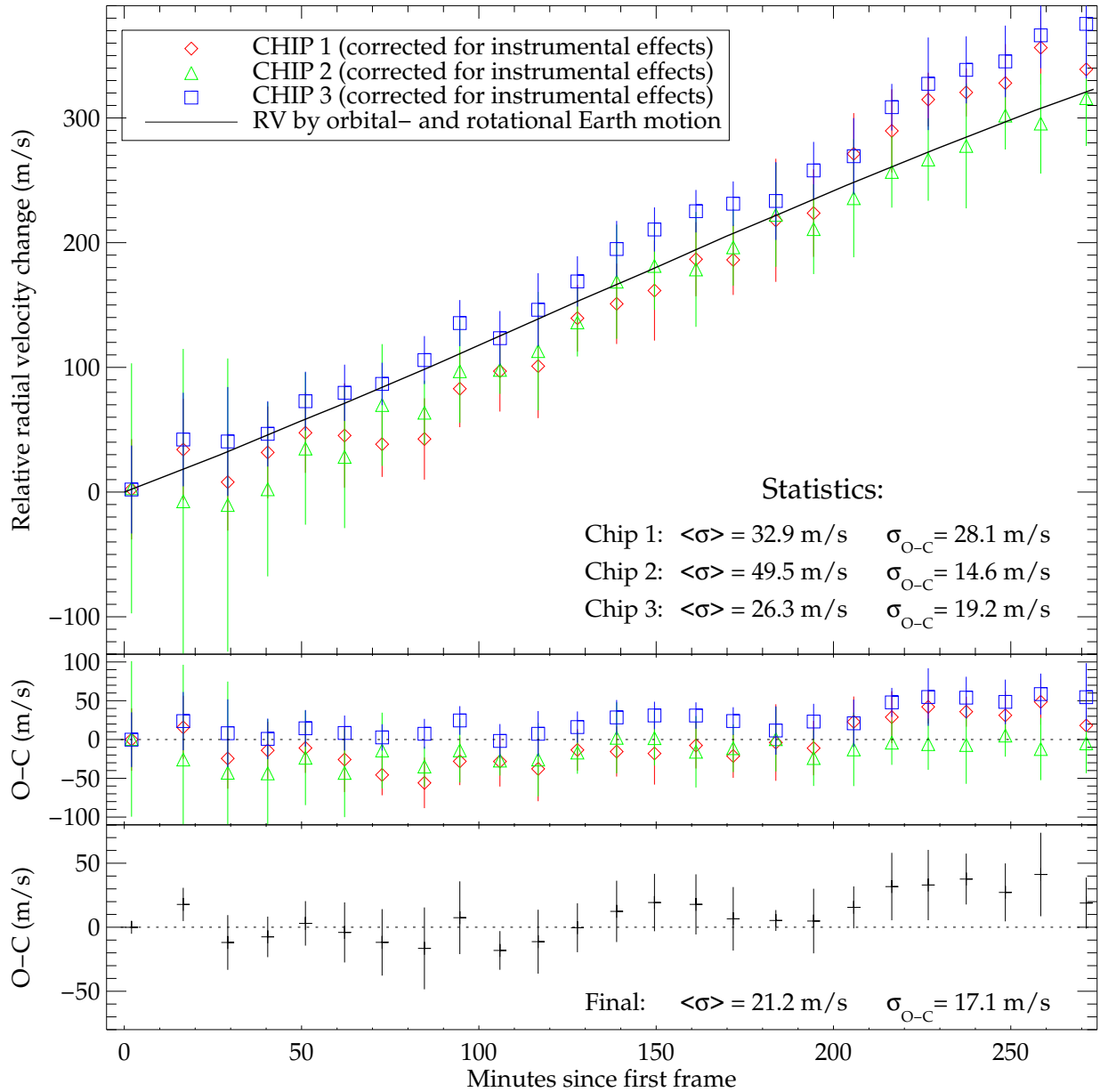}
\caption{Radial velocity of the science target -- MS~Vel -- after subtraction of the radial velocity of telluric absorption lines imprinted in the same spectrum (see Fig.~\ref{fig:sourcespectrum}). The linear trend is modeled by the barycentric and rotational velocity of the Earth during the observation. In the subpanels the data after subtraction of the model is shown
in terms of observed - computed (O-C), first separately for each of the three chips analyzed here, finally for the mean of all three chips. The standard deviation of the mean was used to derive formal error bars. See text for details.
\label{fig:RVsource}}
\end{figure*}

Subtracting the RV of the atmospheric absorption lines (that suffer the same 
displacements) from the RV measured in the masked source spectrum cancels out 
most of the effects of the nodding displacement and the source misalignment due
the refraction correction. 
A linear trend is left, in the counter direction of the trend observed in the 
individual spectra of the source, peaking at 300\,m/s after 4.5~hours. The 
results at this point of the analysis are displayed in 
Fig.~\ref{fig:RVsource}. Again, we combine here the 12 spectra from each of 
the 25 nodding cycles to one single data point. 

The remaining trend can be modeled by the barycentric motion and rotation of the
Earth. Subtracting this trend leaves residuals in the order of approx. $\pm$ 20
m/s. The residuals appear not noise-like, even though the formal uncertainties
of the RV points would account for most of the amplitude of these residuals,
but show a rather smooth and semi-periodic behaviour. These
residuals are not correlated with the residuals in the atmospheric emission
lines (see Fig.~\ref{fig:RVemission}) making it unlikely that the
variability of atmospheric lines , e.g. by wind drifts, are responsible for 
the residuals in the data. It is also very unlikely that the source itself 
shows a RV variability within one or two hours, given the large radius of the 
giant star. Hence, we assign the residuals to a general noise floor limiting 
the RV precision to this level, being most likely due to the data analysis
rather than intrinsic to instrumental instabilities. 

We note, that our analysis does not take changes in the line-profile into
account. This is an essential part of the analysis when aiming for very high
radial velocity precision \citep{Valenti95, Butler96} using in-situ 
measurements through a gascell. The method involves a template spectrum of the
gascell with much higher resolution than the one used at the actual science
measurements. This template spectrum is used to deconvolve the measured lines 
to determine the instrumental profile. We did not dispose of such a spectrum 
of our N$_2$O gascell, nor do we have such a template for the telluric lines. 
Hence, this shortcoming in our analysis can easily account for the remaining 
residuals. Moreover, the masking technique to separate the telluric and stellar
lines in the individual spectra is not perfect for highly blended line systems, 
as is the case here.

\section{Conclusions}
We have performed the first radial velocity stability test of CRIRES with 4.5~h
of on-sky measurements of the giant MS~Vel at 4100\,nm. The dispersive
parts of the spectrograph have been electronically stabilised, thus decoupling
the problems of functional repeatability, spectrograph stability  
and tracking errors. 
Besides this non-standard settings we used the spectrograph as offered to the
community, which demonstrates that even at this very early stage of its life,
CRIRES is a stable instrument, ready to measure radial velocities with high
precision including a potential for even higher performance when using an
appropriate gas as local RV zero point.

We could demonstrate that CRIRES shows only a slow drift in its intrinsic radial 
velocity zero point of $\sim$ 60\,m/s in 4.5\,h. This drift is fully compensatable 
using a N$_2$O gas-cell in calibration mode or telluric lines as a local rest frame.
Moreover, if our assumption holds that the drift is due to a small change in temperature of 
the grating (by approx. 1/20~K), it will not appear in any later measurements
as the grating is now actively stabilised to approx. 1~mK.

The telluric lines in emission, observed with a resolution of R$\sim$50\,000
are stable in RV to a level of $\leq \pm$ 10\,m/s and are therefore suitable as a
local rest frame for high precision RV measurements. For the relative radial 
velocity of our science target, MS~Vel, we could recover the 300\,m/s drift 
induced by rotational and barycentric Earth's motion. The remaining residuals (rms)
in the stellar spectra are $\leq \pm$ 22\,m/s.

We can conclude that the telluric lines imprinted in the science object are
not a limiting factor for precise RV measurements in the near infrared but can
instead be used as a substitute for a gas-cell providing the necessary
rest frame in cases where no suitable gas can be found or the interesting
stellar features are in regions of high telluric line density. First scientific
results using this method are presented in \citet{TWHya}.

The influence of mismatches between the diffraction limited FWHM of the PSF and
the typical slit width in combination with dynamic (i.e. tracking induced)
misalignments of the PSF in respect to the slit centre are substantial. 
However, these problems can be compensated when telluric or gas-cell lines are
measured simultaneously with the source and used as a local rest frame. This
method is well known and widely used in the optical, where Iodine lines are used 
as the local restframe.
Given the high rate of close blends of telluric lines and photospheric lines from
our science target and taken the limitation of our analysis into account, we see 
a clear margin for a gain in performance. Especially the use of a gas-cell, well
characterised by a high resolution laboratory spectrum (R$\gg$100\,000), should
provide a better local rest frame when aiming at RV precision below $\approx$20\,m/s. 
The gas has to exhibit a high and stable line density in regions where the science target
shows also a reach spectrum. To avoid blends with telluric lines, rare isotope 
enhanced gases could be used which would elegantly avoid problems with blending 
telluric features. 

In the example presented here - as for all conceivable RV applications of CRIRES - 
rotational vibrational transitions of molecules in the vibrational ground-state are 
used.  These gases behave very close to ideal gases. This implies that the density 
of molecules in the beam is invariable and relatively easy controllable. This situation 
is fundamentally different in the Iodine gas-cells which are being used for radial 
velocity work. There iodine vapour is used in a heated cell and condensation on 
cooler parts can result in strong changes of the vapour pressure. Therefore, in our 
case it is safe to assume, that the transition frequencies are very close to those, 
which have been determined by NIST using heterodyne-techniques relative to the Caesium 
time standard. The spectra are relatively simple and lines which may consist of a blend 
can be positively excluded. The only conceivable process which may change transition 
frequencies in the gas-cell in the case of an ideal gas is pressure shift. The pressure 
may change due to the external temperature, which will be monitored. It has been estimated 
that this effect is less than ~1m/s. A more serious problem is pressure shift by air, 
in case the gascell has a leak. Fig. 2b in \citet{Glenar} gives an example. A PH3-gascell 
line in the presence of a leak is measured relative to a gain-stabilised CO2-laser in a 
heterodyne setup. The kinks in the PH3-trace correspond to refillings of the cell.
To control this problem, the pressure in the CRIRES gas-cells will be monitored as part of 
regular operations.

A gascell filled at low pressure with species, for which NIST has done a heterodyne 
reference measurement relative to the time standard (Caesium-clock), in principle provides 
for an absolute secondary frequency standard. To that end CRIRES measurements are 
linked in an absolute and traceable way to the time standard. This is, strictly speaking,
not an actual experimental proof, that this indeed allows for a "cm/s"-type calibration of 
the spectrograph. Indeed e.g. a spurious blend of gas-cell lines with weak telluric lines  
may produce artificial shifts exceeding this value. Experts in the field indeed stress that 
the frequency calibration of spectrographs breaks down at a precision of $10^{10}$ and only 
heterodyne techniques can provide for this precision (Theodor Haensch, priv. communication).

We finally note that this paper aimed at demonstrating the short-term stability 
of CRIRES and the implications when observing with an underfilled entrance slit. The long-term 
stability of the instrument, especially important for RV studies, will be considered 
in a later paper.

\begin{acknowledgements}
We are grateful to all colleagues involved in the development, installation and
commissioning of CRIRES. AS acknowledges financial support from the 
Deutsche Forschungsgemeinschaft under DFG RE 1664/4-1.
  \end{acknowledgements}

\clearpage

\onecolumn

\end{document}